# Stars Approaching the Substellar Limit in the $\alpha$ Persei Open Cluster. [*]

**M.R. Zapatero Osorio**[1], **R. Rebolo**[1], **E.L. Martín**[1], and **R.J. García López**[1,2]

[1] Instituto de Astrofísica de Canarias, Vía Láctea s/n, E-38200 La Laguna, Tenerife, Spain
[2] Department of Astronomy, The University of Texas at Austin, RLM 15.308, Austin, Texas 78712-1083, USA

**Abstract.** We present intermediate dispersion optical spectroscopy for seven very low-mass stars in the $\alpha$ Per open cluster with spectral types between M3 and M6, including the brown dwarf candidate of Rebolo et al. (1992). Our radial velocity measurements are found to be generally consistent with the mean cluster velocity to within the measurement errors. H$\alpha$ equivalent widths have been measured and compared to previous published data for other known members of the cluster. A turnover in chromospheric activity around spectral type M3–M4 is observed. The brown dwarf candidate, located in the cool side of the turnover, is confirmed to exhibit a strong H$\alpha$ variability. It is the coolest $\alpha$ Per candidate member for which spectra are available and for which chromospheric activity has been measured.

Using the Li I doublet at $\lambda 6707.8$ Å, we derive upper limits to the atmospheric Li abundance in the sample and discuss them in the context of the most recent stellar evolutionary tracks. The non-detection of the Li I line in the spectrum of the brown dwarf candidate implies a mass greater than 0.08 $M_\odot$, and therefore it is not a substellar object. However, its position in the HR diagram implies that this star is indeed very close to the substellar limit.

## 1. Introduction

Considerable attention in searches for low-mass stars has been devoted to the $\alpha$ Per open cluster (Prosser 1994a, 1992; Stauffer et al. 1989, 1985). With an age estimated from the main-sequence turn-off of 50 Myr (Mermilliod 1981) and metallicity close to solar, $\alpha$ Per is one of the nearest clusters to the Sun ($d \simeq 170$ pc, (m-M)$_\circ$ = 6.1, Stauffer et al. 1989; Crawford & Barnes 1974). Its young age and proximity make $\alpha$ Per one of the most privileged places to carry out detailed spectroscopic studies of low mass stars. Observations and subsequent analysis of emission lines and light element depletion in the stellar members of the cluster will provide significant insights on the physics of their interiors, the structure of the outer stellar layers, mixing processes and convection.

It is already well established that the vast majority of young very low mass (hereafter referred to as VLM) stars show H$\alpha$ and the Ca II resonance lines in emission (e.g. Cram & Giampapa 1987). In fact, these emissions have been used as a criterion for membership to young clusters (Prosser, Stauffer & Kraft 1991, and references therein). Recently, Stauffer et al. (1994) have measured H$\alpha$ emission in VLM members of the Hyades and Pleiades clusters finding an "apparent break" in the chromospheric activity in the Pleiades at around (V-I$_K$) $\sim$ 3.0, while not in the Hyades. It is worth looking for a similar feature in the $\alpha$ Per open cluster as it is very close in age and metallicity to the Pleiades and, therefore, some similarities can be expected.

Observations of Li in late-type stars that are either still approaching the main sequence or very recently settled on it are crucial for the knowledge of the pre-main sequence Li depletion and the mechanisms leading to its destruction. Li abundances in K-type stars of $\alpha$ Per (Balachandran, Lambert & Stauffer 1988) show that a large depletion has already taken place at the age of the cluster. García López, Rebolo & Martín (1994) conclude that the destruction factor is larger than 1000 for early-M stars in $\alpha$ Per. It yet remains empirically unknown what may occur at later spectral types. According to theoretical calculations, stars with masses greater than 0.08 $M_\odot$ should have destroyed Li at the age of 50 Myr, while less massive objects, i.e. brown dwarfs, should preserve it (e.g. Magazzù, Martín & Rebolo 1993; D'Antona & Mazzitelli 1994). Rebolo, Martín & Magazzù (1992) (see also Magazzù et al. 1993) have proposed that the detection of Li I in VLM objects is feasible and can provide

**Table 1.** Log of spectroscopic observations

| Name | Telescope | Date obs. (UT) | $t_{exp}$ (s) | Sp. range (Å) | Disp. (Å/pix) |
|---|---|---|---|---|---|
| AP J0323+4853* | 4.2m WHT | Jan 22, 1993 | 1800 | 5500–6950 | 1.47 |
|  | 3.5m Calar Alto | Aug 10, 1993 | 1800 | 6500–7200 | 0.89 |
|  | 3.5m Calar Alto | Aug 12, 1993 | 1800 | 6500–7200 | 0.89 |
|  | 4.2m WHT | Sep 16, 1993 | 1200 | 6100–9000 | 2.72 |
|  | 4.2m WHT | Dec 17, 1993 | 2x2200 | 6100–6900 | 0.74 |
| AP176 | 3.5m Calar Alto | Aug 7, 1993 | 2x1800 | 6500–7200 | 0.89 |
| AP246 | 3.5m Calar Alto | Aug 9, 1993 | 900 | 6500–7200 | 0.89 |
| AP237 | 3.5m Calar Alto | Aug 9, 1993 | 2x1800 | 6500–7200 | 0.89 |
| AP179 | 2.5m INT | Oct 26, 1993 | 3x1200 | 6400–7700 | 1.12 |
| AP143 | 2.5m INT | Oct 27, 1993 | 1000 | 6400–7700 | 1.12 |
| AP240 | 4.2m WHT | Dec 19, 1993 | 1800 | 6100–6900 | 0.74 |

* Other name: AP275 (Prosser 1994a).

a powerful spectroscopic diagnostic of their nature. Stars of spectral types M3–M4 in $\alpha$ Per are believed to have $\sim$ 0.3–0.2 $M_\odot$. The latest M-type objects of the cluster, like the M6 dwarf AP J0323+4853 [1] (Rebolo et al. 1992), are located in the HR diagram in positions that suggest masses about 0.08 $M_\odot$ and even lower, therefore Li may be present in their atmospheres and could be detected. Both, detection or non-detection of Li would provide significant insights on the mixing processes in the convection zone of VLM stars and on the location of the substellar limit in the $\alpha$ Per open cluster.

In this paper, we present results both on the H$\alpha$ emission and on the Li resonance doublet at $\lambda$6707.8 Å among very cool $\alpha$ Per stars (M3–M6), and compare our observational results to theoretical predictions of Li depletion within the context of new VLM stars and brown dwarfs evolutionary models. In Section 2 we discuss the observations; in Section 3 we present the analysis of the spectroscopic data (radial velocities and H$\alpha$ equivalent widths); in Section 4 we derive upper limits to the Li abundances; and in Section 5 conclusions are presented.

## 2. Observations

The sample consists of six late-type stars (AP143, AP176, AP179, AP237, AP240, AP246) taken from Prosser (1992). This author classified them as members of the $\alpha$ Per cluster as a result of combined astrometric, photometric and low resolution spectroscopic observations. Also included in the sample is AP J0323+4853, the brown dwarf candidate of Rebolo et al. (1992). Our radial velocity measurements support the membership of this object which has the latest observed spectral type ($\sim$M6) of any known member.

[1] We note that this object was originally named in Rebolo et al.(1992) as Ap 0323+4853. According to IAU prescriptions on nomenclature of new objects with numbering based on 2000 coordinates, they should be preceded by J. In addition, to avoid confusion with the Ap spectral classification we have capitalized the acronym. Therefore, the new designation that will be used throughout this paper is AP J0323+4853.

Intermediate resolution spectra for the seven late-type stars have been obtained using the 4.2 m WHT and the 2.5 m INT at the Observatorio del Roque de los Muchachos (ORM) located on the island of La Palma (Canary Islands, Spain), and the 3.5 m telescope at Calar Alto (Almería, Spain). The instrumentation used was the ISIS double arm spectrograph on the WHT, the IDS spectrograph on the INT and the TWIN spectrograph on the 3.5 m telescope. In Table 1 we list by chronological order the stars for which spectra have been taken, the observing date, the exposure time, the spectral coverage and the nominal dispersion obtained with each instrument. The slit projection was typically two pixels.

Each individual spectrum was reduced by a standard procedure using various tasks within the IRAF [2] package, which included debias, flat field, optimal extraction and wavelength calibration using arc lamps. The Calar Alto spectra were flux calibrated using the standard star HD19445 which has absolute flux data files available in the IRAF environment. INT and WHT spectra were not flux calibrated as no standards were observed during the nights. In Fig. 1 we present the final spectra of the seven programme stars. The relatively high S/N (20–35) spectra of AP179, AP237, AP246 and AP240 illustrate the real features present in these wavelength regions for stars of spectral type M3–M5. The strongest features are TiO bandheads and continuum peaks plus sharp H$\alpha$ emission lines (see Kirkpatrick, Henry & McCarthy 1991).

Prosser (1994b) has very kindly provided us with several spectra of M4–M5 type stars that are also believed to be candidate cluster members on the basis of their photometry and H$\alpha$ emission. These are AP268, AP272, AP279, AP284 and AP296 (see Prosser 1994a). AP272 and AP279 are the reddest objects for which Prosser (1994a) has published spectra.

[2] IRAF is distributed by National Optical Astronomy Observatories, which is operated by the Association of Universities for Research in Astronomy, Inc., under contract with the National Science Foundation.

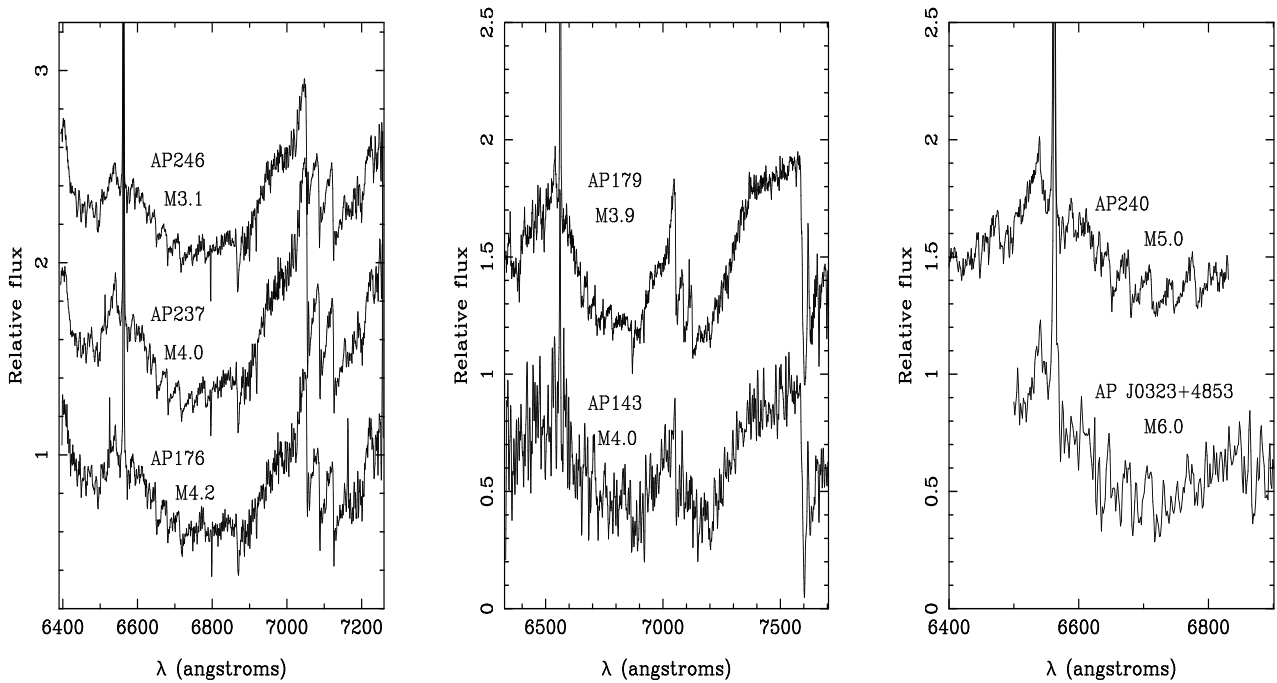

**Fig. 1.** Final spectra of our sample of seven stars

## 3. Analysis

*3.1. Spectral types, luminosities and effective temperatures*

Some helpful information for our sample and for the stars provided by Prosser is listed in Table 2. Spectral types, V magnitudes and (V−$I_K$) colours were taken from Prosser (1992, 1994a) except for AP143 and AP J0323+4853 for which we have derived their spectral types using the A index defined by Kirkpatrick et al. (1991). This is a reliable spectral type index because it shows a strong dependence with spectral type and the scatter for each individual M subclass is small as compared to other calibrations. AP J0323+4853 is M6 and AP143 is M4. The uncertainty in these estimates is about half a subclass. In the spectra of these two stars lines of K I ($\lambda\lambda$7665, 7699 Å) are observed (in AP J0323+4853 is also obvious the Na I $\lambda\lambda$8183, 8195 Å doublet as the spectral coverage is larger). According to several luminosity discriminants described in Kirkpatrick et al. (1991), these two stars are dwarfs. In order to check the consistency between Prosser's spectral types and those derived using the A index of Kirkpatrick et al. (1991), we have also determined the spectral types for the rest of the stars in the sample finding good agreement.

The luminosity of our objects was determined as the average value of those obtained using four different calibrations: Bessell & Stringfellow (1993) and Bessell (1991) in terms of the bolometric correction in I filter as a function of (V–$I_C$), and relations between bolometric magnitudes and absolute $M_V$ and $M_I$ magnitudes given in Bessell & Stringfellow (1993). The final luminosity values and the standard deviations with respect to the mean are given in Table 2. Effective temperatures for our stars were obtained using the $T_{\rm eff}$ versus (V–I) calibration from Kirkpatrick et al. (1993), slightly extrapolated in the case of AP246. The (V–$I_K$) colours in Table 1 were translated into the Cousins system using the relations derived by Bessell & Weis (1987). At present, there is no reliable $T_{\rm eff}$ calibration for stars cooler than 4000 K. The discrepancies when considering different authors can be as high as 500 K (see Martín, Rebolo & Magazzù 1994a), much larger than the estimated internal uncertainties (±100 K). Using Kirkpatrick et al.'s (1993) calibration we plot our stars in the HR diagram of Fig. 2. A discussion follows below on the difficulties in deriving masses due to different $T_{\rm eff}$ calibrations and to different theoretical models.

In Fig. 3.1 D'Antona & Mazzitelli's (1994) evolutionary tracks (masses ranging from 0.4 down to 0.04 $M_\odot$) and isochrones are plotted. These authors give pre-main sequence evolutionary models for a wide range of masses (down to 0.02 $M_\odot$). We chose those that assume the convection treatment proposed by Canuto & Mazzitelli (1991) and the opacity tables of Rogers & Iglesias (1992) and Alexander et al. (1989) because the authors claim that this computed set is the best to describe the location of the low-mass sequence in a luminosity-$T_{\rm eff}$ diagram.

The location of our stars in the HR diagram, when compared with D'Antona & Mazzitelli models, adopting either the $T_{\rm eff}$ calibration given in Kirkpatrick et al. (1993) or that of Bessell (1991) suggests a very young age for $\alpha$ Per (younger than

**Table 2.** Photometric and spectroscopic data

| Name | V | (V-I$_K$) | SpT | $T_{\rm eff}$ $\pm 100$ (K) | log ($L/L_\odot$) | $M/M_\odot$ $\pm 30\%$ | $v_r^\dagger$ (km/s) | EW(H$\alpha$) $\pm$ 0.5 (Å) | EW(Li I) (mÅ) | log N(Li) (NETL) |
|---|---|---|---|---|---|---|---|---|---|---|
| AP246 | 15.24 | 2.41 | M3.1 | 3640 | -0.95±0.14 | 0.40 | -13.7 | 4.2 | <130 | <0.11 |
| AP176 | 16.41 | 2.70 | M4.2 | 3490 | -1.26±0.12 | 0.30 | -0.1 | 6.7 | <150 | <0.22 |
| AP179 | 16.77 | 2.74 | M3.9 | 3470 | -1.37±0.10 | 0.25 | -10.5 | 9.3–10.8 | <100 | <-0.04 |
| AP237 | 16.84 | 2.90 | M4.0 | 3400 | -1.34±0.14 | 0.20 | 9.8 | 7.5 | <115 | <0.05 |
| AP143 | 18.00 | 2.92 | M4.0 | 3390 | -1.73±0.06 | 0.20 | -15.7 | 3.6 | <525 | <1.90 |
| AP240 | 17.22 | 3.15 | M5.0 | 3300 | -1.39±0.17 | 0.15 | -1.0 | 7.2 | <110 | <0.01 |
| AP296 | 18.57 | 3.22 | M4.7 | 3270 | -1.83±0.08 | 0.15 | | 6.2* | <135 | <0.16 |
| AP284 | 19.44 | 3.38 | M4.6 | 3210 | -2.07±0.05 | 0.10 | | 5.4* | <270 | <0.23 |
| AP268 | 20.50 | 3.61 | M5.2 | 3140 | -2.36±0.05 | 0.150–0.040 | | 4.2* | <405 | <0.81 |
| AP272 | 20.42 | 3.77 | M5.6 | 3080 | -2.28±0.07 | 0.100–0.035 | | 13.5* | <885 | <1.94 |
| AP279 | 21.10 | 3.86 | M5.0 | 3050 | -2.49±0.05 | 0.095–0.030 | | 5.2* | <600 | <1.39 |
| AP J0323+4853 | 21.10 | 3.95 | M6.0 | 3040 | -2.46±0.06 | 0.090–0.030 | -5.9 | 4.5–12.1 | <450 | <0.94 |

NOTES.
* H$\alpha$ EW for AP296, AP284, AP268, AP272 and AP279 are taken from Prosser (1994a). No radial velocities have been derived for these stars because of uncertainties in wavelength calibration.
$\dagger$ For uncertainties in radial velocities see section 3.2.

50 Myr), which does not agree with previous estimates (50 Myr, Mermilliod 1981; Stauffer et al. 1985; 70 Myr, Prosser 1992). In Fig. 3.2 we show the comparison with Burrows et al. (1993) evolutionary tracks (0.1, 0.09, 0.08 and 0.06 $M_\odot$) and isochrones. Here, if we adopt Kirkpatrick et al.'s (1993) temperature calibration, we infer an age closer to that estimated from the main-sequence turn-off (i.e. 50 Myr), but we have to rely in the comparison with only a few objects at the bottom of the diagram.

The differences between the two sets of models are remarkable and give a clear indication of the uncertainties involved in the estimates of masses at these very young ages. As an example, we note that while D'Antona & Mazzitelli (1994) yield a mass for our lowest luminosity object, AP J0323+4853, of 0.052±0.010 $M_\odot$ (assuming Kirkpatrick et al's $T_{\rm eff}$ calibration), well below the substellar limit, Burrows et al. (1993) place this star out of the substellar region with a mass of 0.09±0.01 $M_\odot$. From inspection of Figs. 3.1 and 3.2, we see that the mass uncertainties as a consequence of only differences in temperature calibrations are $\sim$ 0.02–0.04 $M_\odot$ for objects in a range of 0.1–0.05 $M_\odot$. When the differences in the models are also considered we obtain for each of the least massive objects the mass range listed in Table 2. They imply log $g$ values in the range 4.6-5.0. In section 4, we will analyze the consistency of these mass ranges with the limits imposed by the observed depletion of Li.

### 3.2. Radial velocities

All our spectra include H$\alpha$, which is seen in emission. The procedure to derive the radial velocity was the same than in Martín et al. (1994a); first, all the spectra were shifted to the same zero point using ten sky lines and after this, the radial velocities were corrected from diurnal, barycentric and annual

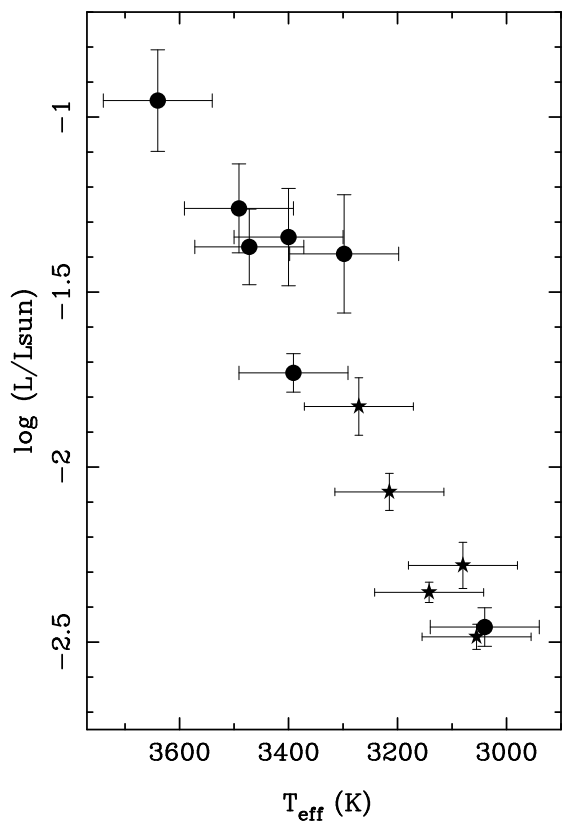

Figure 2.

**Fig. 2.** Locations in the HR diagram of our programme stars. Filled circles stand for our sample (Table 1) while filled stars stand for several very red stars from Prosser (1994a)

Figure 3.1 Figure 3.2

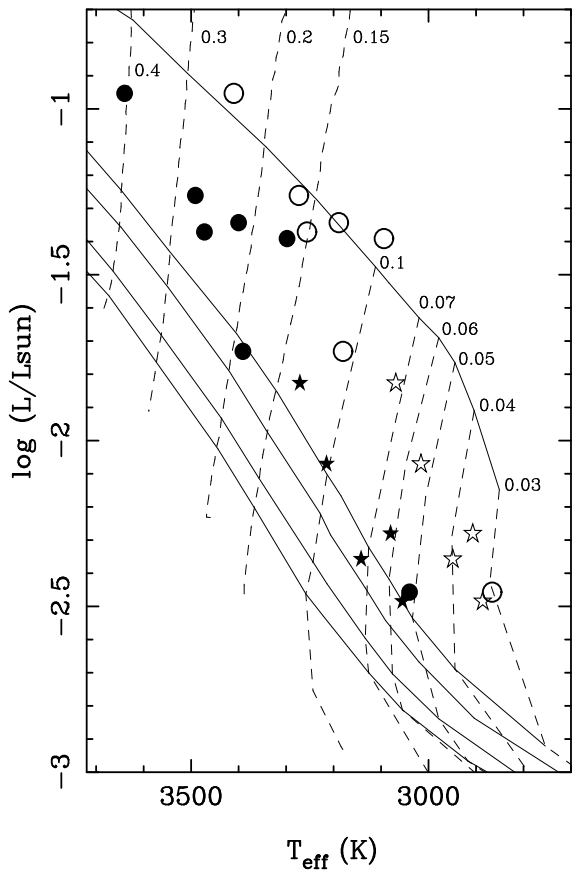 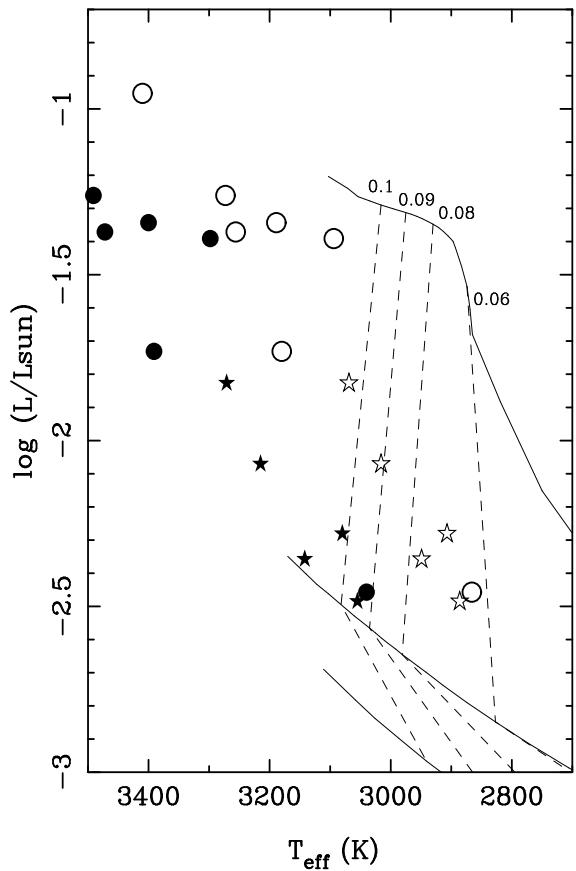

**Fig. 3.** HR diagram overplotting theoretical pre-main sequence evolutionary models of D'Antona & Mazzitelli (1994) [Alexander + Rogers & Iglesias opacities, CM convection, Y=0.28, Z=0.019] (Fig. 3.1) and Burrows et al. (1993) (Fig. 3.2). Symbols and error bars are as in Fig. 2 with an additional information: filled symbols stand for Kirkpatrick et al. (1993) $T_{eff}$ calibration while open symbols stand for Bessell (1991) $T_{eff}$ calibration. Evolutionary tracks are labelled with masses in solar units. In Fig. 3.1, from top to bottom, isochrones are for 3, 20, 30, 50 and 70 Myr. In Fig. 3.2 isochrones are for ages of 3, 70 and 600 Myr

velocities to yield heliocentric radial velocities (listed in Table 2). We measured the centroid of H$\alpha$ and taking into account the dispersion of our spectra, slit projection and accuracy of the determination of the centroid, we estimate that the 1$\sigma$ uncertainties in each individual measured radial velocity are typically in the range 5-10 km/s. We have five independent radial velocity measurements for AP J0323+4853 and the derived mean value with its error is $v_r = -5.9 \pm 2.5$ km/s. The large H$\alpha$ variability found in AP J0323+4853 (see section 3.3.1), seems to have no relation with changes in radial velocity as deduced from the relative small standard deviation of the measurements.

As a check for our radial velocity determinations, during the run that took place at the WHT on December the K-type star AP264 (which has H$\alpha$ in emission, EW = 0.2$\pm$0.5 Å) was also observed. Using the technique described above we derived +2.1$\pm$1.6 km/s, a value which is in very good agreement with that obtained by Prosser (1992) (+2.4$\pm$1.2 km/s). With a mean cluster velocity of –2 km/s for $\alpha$ Per, the radial velocities of our sample of stars are consistent within the error bars and provide reinforcing evidence for cluster membership.

### 3.3. H$\alpha$ equivalent widths and chromospheric activity on VLM stars in $\alpha$ Per

Our H$\alpha$ equivalent widths (hereafter referred to as EW) are listed in Table 2. Typical errors are $\pm 0.5$ Å except for the low S/N ratio spectra (AP143 and AP J0323+4853) for which they are as large as twice, due to the difficulty in continuum placement. We have checked that our location of the continuum is consistent with the one used in Prosser (1992, 1994a) by finding a good agreement between our measurements on the spectra he provided us and his published EWs. Figure 4 shows the H$\alpha$ EW as a function of the (V$-$I$_K$)$_\circ$ unreddened colour for K and M dwarfs (assuming a mean reddening of E(V$-$I$_K$)=0.16).

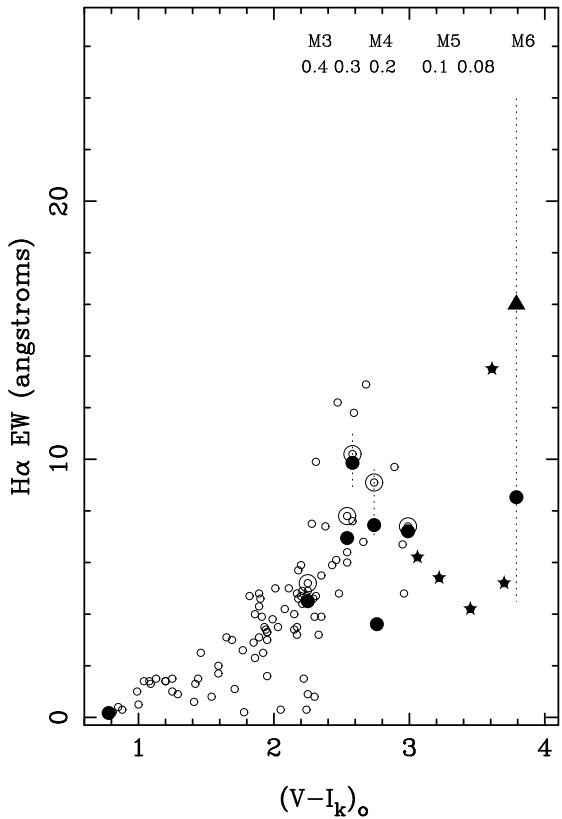

**Fig. 4.** Hα equivalent width as a function of (V-I$_K$) dereddened colour for α Per. Filled symbols: the programme stars (as in Fig. 2). For AP J0323+4853 and AP179 the mean values of the Hα EW's measured in this work are plotted. Open circles: stars with Hα in emission taken from Prosser (1992, 1994a). Double circles: stars from Prosser (1992) in common with ours. Filled triangle: mean value of the data for AP J0323+4853 from Rebolo et al. (1992). The hottest star is AP264 (see Section 3.2). Stars with Hα variability are plotted with a dotted line indicating the range of the measurements. The largest variability corresponds to AP J0323+4853. On the top, spectral types and approximate masses in solar units are indicated (assuming 50 Myr as the age of the cluster and D'Antona & Mazzitelli 1994 theoretical evolutionary tracks)

### 3.3.1. Hα variability

While this paper is not mainly aimed at the study of Hα chromospheric emission, we have observed some peculiarities which should be noted. Our Hα EWs for AP176, AP240 and AP246 agree within the error bars with those from Prosser (1992). For AP179 and AP237 we find some indication of variability, at a level 20% that slightly exceeds the uncertainties of the measurements. AP143 appears to be out-of-place in Fig. 4. It may be due to uncertainties in reddening correction, since it is located in an area for which the colour excess is about 0.16–0.30 mag. However, our M4 spectral type classification is consistent to obtain additional spectroscopy to see if there is indication of variability in this star.

**AP J0323+4853.** We confirm a strong Hα variability in AP J0323+4853, already reported by Rebolo et al. (1992), who measured significant changes in the emission in intervals of half an hour. We have also found variability at a level 45% within approximately the same interval of time. In Table 2 we list the Hα EW range measured in the present work. Only the mean values for the two epochs of observations are plotted in Fig. 4, while the whole range of variability is indicated by a vertical dotted line.

This Hα behaviour could be explained in different ways: (i) AP J0323+4853 may be a slow flare star which shows a brightness decrease after a maximum similar in each flare (we note that in any consecutive observation, with time separation of typically half an hour, Hα has always been observed as a progressive decreasing emission), or (ii) there may be some chromospheric local structures such as extensive regions of strong excitation, and in such case a periodicity should be expected due to the axial rotation of the star, or (iii) simply a combination of both hypothesis. If variability in Hα is somehow related to rotation then the monitoring of this emission may yield rotational velocities and periods, quite unknown parameters for late-M stars. Zapatero Osorio, Rebolo & Martín (1995) are monitoring some of these stars (including AP J0323+4853) for rotational modulation. Our preliminary results suggest a rotational period for this object that is consistent with the scale of variability in Hα.

### 3.3.2. Hα $vs$ mass

A peculiar feature in Fig. 4 is the apparent $turnover$ in chromospheric activity in α Per near (V−I$_K$)$_\circ$ = 2.9. Stars redder than this value show a lower level of Hα emission than stars just blueward that colour. In the Pleiades, a similar behaviour has been detected by Stauffer et al. (1994), and more recently confirmed by Hodgkin, Jameson & Steele (1995) and by Stauffer, Liebert & Giampapa (1995), at approximately the same colour. This is not the case for the Hyades, a cluster nearly ten times older than the previous two. Hodgkin et al. (1995) have also studied the X-ray coronal activity for a large number of low and very low-mass Pleiades stellar members. They find that in both cases, the subsequent activity decreases for stars with M$_I$ > 8–9 ((V-I$_K$)$_\circ$ > 3), and suggest that the observed turnovers are due to the transition from radiative to convective cores. By inspecting D'Antona & Mazzitelli (1994) pre-main sequence evolutionary models, we find that at masses about 0.3–0.2 $M_\odot$ this transition theoretically takes place regardless of age. From Fig. 4 we observe that the turnover occurs just for this range of masses suggesting that theory and observations may agree on the "discontinuity" in Hα emission being caused by the appearance of a fully convective state in these stars, and hence, a change in the nature of the magnetic dynamo must be expected. However, the detection of high levels of Hα emission in the fully convective object AP J0323+4853, implies that there can

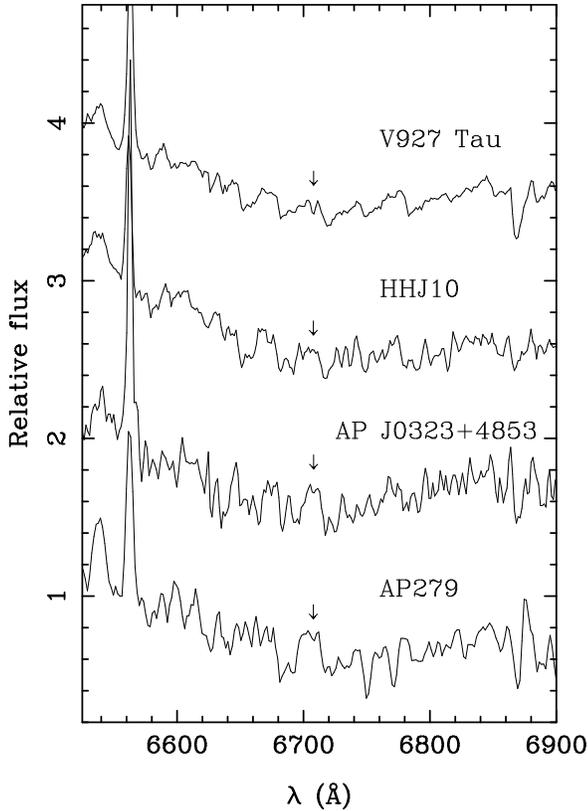

**Fig. 5.** Spectra of very cool young objects in the Li I λ6707.8 region. The location of the Li I line is indicated by an arrow in each spectrum

## 4. Li and very low mass stars in α Per

We have found no evidence for Li in any of the spectra. The $1\sigma$ upper limits to the Li I EWs, given in Table 2, were derived considering the strongest possible feature that could be present in the region around λ6707.8 Å and taking into account the S/N ratio and resolution of each spectrum.

In Fig. 5 we plot the spectra of AP J0323+4853, AP279, HHJ10 (one of the faintest and reddest proper motion members of the Pleiades; Hambly, Hawkins & Jameson 1993) and V927 Tau (a M5 T Tauri star). The last two spectra, taken from Martín et al. (1994a), are included for comparison. Only the T Tauri star shows Li in the spectrum. The weak features at λ6707.8 Å in AP J0323+4853 and AP279, although suggestive, are below our detection level and only upper limits to the Li EW can be established. On the other hand, the TiO molecular bands at λ6569, λ6630, λ6650, λ6680, λ6713 and λ6852 Å can be identified in the spectra.

to the Li abundance was established for each star using the NLTE curves of growth computed in Pavlenko et al. (1995). These curves are obtained assuming a Li atom model of 20 levels and Kurucz (1992) model atmospheres. Our upper limits are listed in Table 2 in the standard scale in which log N(H) = 12 and shown as a function of log $T_{\rm eff}$ in Fig. 6. A large Li depletion, by a factor $\sim$ 1000, is found in the hottest programme stars with $T_{\rm eff} > 3150$ K (log $T_{\rm eff} > 3.498$) (the spectrum of AP143 is noisy and therefore the upper limit is less restrictive). These depletions, similar to those derived by García López et al. (1994) for three early M-type α Per stars, confirm that a very efficient mechanism of Li destruction takes place in the early evolution of VLM stars.

The Li depletion computed by D'Antona & Mazzitelli (1994) for ages ranging from 20 to 100 Myr are plotted in Fig. 6. At the age of 20 Myr, stars with $T_{\rm eff}$'s greater than 3400 K (i.e. masses greater than 0.2 $M_\odot$) are expected to deplete all their initial Li content, while cooler stars are not. The Li observations in our coolest stars are clearly inconsistent with such a short age for the α Per cluster. While the star locations in the HR diagram indicates an age younger than 50 Myr, the Li upper limits strongly suggest an older age. For the least massive objects in our sample the disagreement is even more serious. Objects like AP J0323+4853, according to D'Antona and Mazzitelli tracks should have masses lower than 0.08 $M_\odot$ (i.e. substellar). All theoretical calculations show that below this mass and at ages shorter than 100 Myr the initial Li content is preserved (D'Antona & Mazzitelli 1994; Magazzù et al. 1993; Nelson, Rappaport & Chiang 1993; Bessel & Stringfellow 1993). The Li upper limit in AP J0323+4853 therefore implies a mass greater than 0.08 $M_\odot$, inconsistent with that estimated from the tracks. These discrepancies may be a consequence of assigning too low $T_{\rm eff}$'s for VLM stars, and/or that the tracks predict too hot temperatures (see e.g. Martín et al. 1994a). It is also worth recalling that the HR diagram location of tracks is largely model-dependent. Different assumptions on opacities, equation of state and convection treatment can produce significant effects. The location of AP J0323+4853 in the HR diagram compared with the isochrones and evolutionary tracks taken from Burrows et al. (1993) and Kirkpatrick et al. (1993) $T_{\rm eff}$ calibration yields a mass of 0.09±0.01 $M_\odot$, which is perfectly consistent with the result of the Li test if α Per is just as old as 50–80 Myr. This seems to favour Burrows et al. models versus those by D'Antona and Mazzitelli.

As a summary of our present knowledge about Li abundance and depletion in the α Per cluster, we have compiled data for F, G and K dwarfs from the work of Balachandran et al. (1988) and illustrated them as a shadowed region in Fig. 6. The new corrections from Balachandran (1995) are incorporated in the plot, reducing considerably the scatter in the Li determinations at G and K spectral types. As it can be seen from Fig. 6 the theoretical predictions give, in general terms, a reasonable explanation for the observational features in the diagram. The absence of significant depletion in the more massive stars ($\geq$1.1 $M_\odot$) contrasts with the drastic absence of Li in the VLM stars

Figure 6.

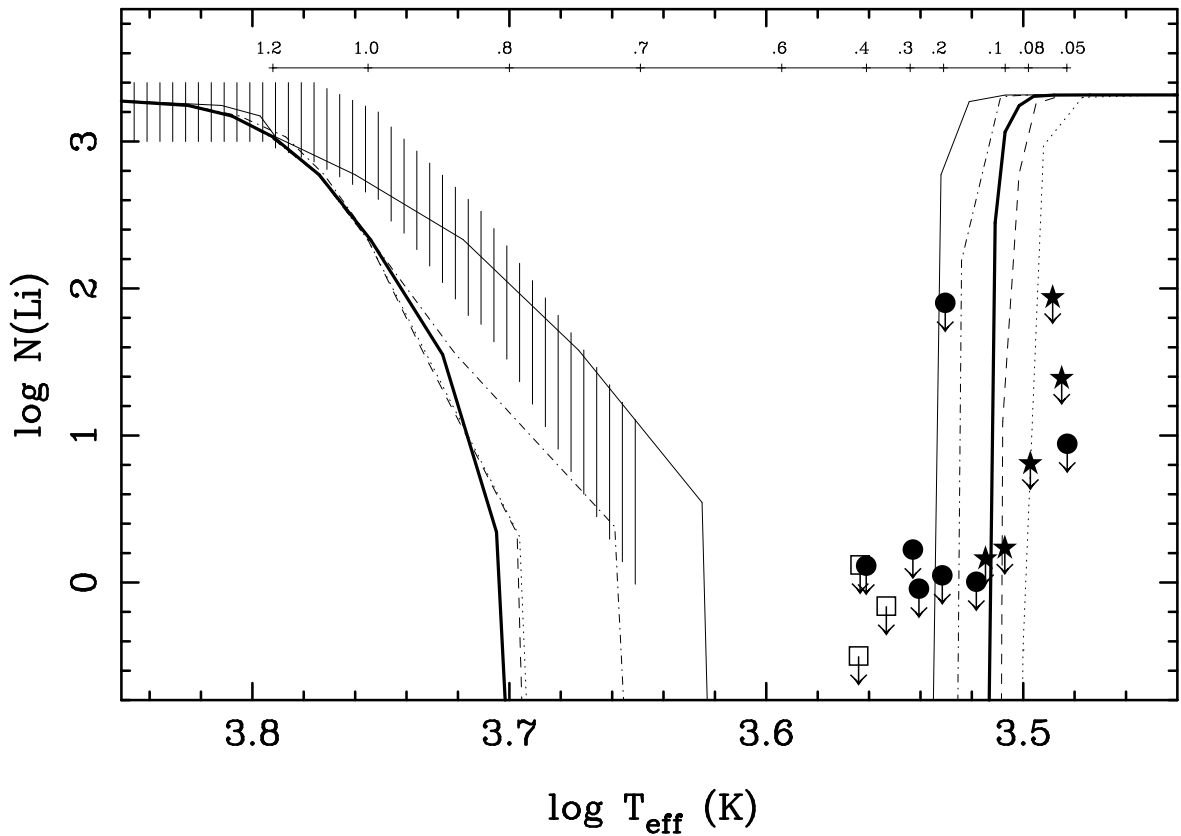

**Fig. 6.** Lithium abundance versus log $T_{eff}$ for stars in $\alpha$ Per. Filled symbols: the programme objects as in Fig. 2. Open squares: stars taken from García López et al. (1994). Shadowed region: location of Li abundances for hotter stars (Balachandran 1988, 1995). For comparison, theoretical Li depletions taken from D'Antona & Mazzitelli (1994) [Alexander + Rogers & Iglesias opacities, CM convection, Y=0.28, Z=0.019] are superposed: thin full line stands for 20 Myr; dash-dotted line for 30 Myr; thick full line for 50 Myr; dashed line for 70 Myr; and dotted line for 100 Myr. On the top, masses in solar units are indicated (as in Fig. 4)

(0.4–0.1 $M_\odot$). Both facts are successfully explained by the theoretical models. However, the slope of the progressive decrease in Li abundance from G towards K spectral types, which sets strong constraints to the modelling of convection in low mass stars, is not so well described by the models. D'Antona & Mazzitelli (1994) predict that at the age of 50 Myr stars with $T_{eff}$ < 5000 K should have depleted their initial Li content by several orders of magnitude while the observations show depletions by only a factor of 10 (i.e. although depletion has already started, it has not become strong yet). Better opacity determinations and a deeper study of the convection problem in these stars is required in order to reach a consensus between observational data and theoretical predictions. Rotation is another parameter that modifies the pre-main sequence tracks and the Li depletion rates (see Martín & Claret 1995), therefore, it should be taken into account when computing models. For very cool temperatures, the models predict an undepleted plateau in Li abundance.

Our work provides evidence that the fainter stars in the cluster, spectral types around M6, still show Li depletions. The search for new $\alpha$ Per members of later spectral types and the observation of Li in them could provide a splendid confirmation of this "cool Li plateau", and more remarkably could establish the existence of substellar objects.

## 5. Conclusions

By inspecting intermediate resolution (1.5–3.5 Å) optical spectra of seven M3–M6 stars in $\alpha$ Per, we have analyzed the H$\alpha$ chromospheric emission and the Li depletion among the less massive stars of the cluster. The late spectral types of the seven stars in our sample, the observed H$\alpha$ lines in emission and the measured radial velocities support their membership to $\alpha$ Per.

In a plot of H$\alpha$ EW versus (V-I$_K$)$_\circ$, a turnover in chromospheric emission is observed at around M3–M4 ($\sim$ 0.3–0.2

dict that they become fully convective. The same behaviour was previously found in the Pleiades by Stauffer et al. (1994) and studied to some extent by Hodgkin et al. (1995) and by Stauffer et al. (1995). AP J0323+4853, which is M6 and therefore placed at the cool side of the turnover, is confirmed to show a strong H$\alpha$ variability indicating rapid changes in the ionization state of the chromosphere and mechanisms that can still support very active chromospheres for fully convective stars.

Li has been largely depleted in M3–M6 stars in $\alpha$ Per. We estimate a minimum Li depletion of a factor of 1000 for M3–M4 type-stars and of 10–100 for M5–M6 type-stars. The efficient Li burning observed implies that the substellar region in $\alpha$ Per has not yet been reached. AP J0323+4853, the faintest and reddest star in our sample, has a mass greater than 0.08 $M_\odot$. This star is the coolest $\alpha$ Per member for which spectra are available and chromospheric activity has been obtained. From the location of this star in the HR diagram a mass in the range 0.09–0.03 $M_\odot$ is obtained, depending on which theoretical models and $T_{\rm eff}$ calibrations are used. Only the uppermost mass in this range is consistent with the observed Li depletion.

*Acknowledgements.* It is a pleasure to thank Dr. Carlos Gutiérrez for his help during the observing run at Calar Alto. We are also indebted to Charles Prosser for kindly providing several of his $\alpha$ Per spectra and for his valuable comments as referee. We thank Dr. M.C. Cortet for indicating us the IAU rules for naming new objects. This research has been partially supported by the Spanish DGICYT project no. PB92–0434–C02.